\documentstyle[12pt]{article}

\newcounter{eq}
\newcounter{sc}

\def\overleftrightarrow#1{\vbox{\ialign{##\crcr
 $\leftrightarrow$\crcr\noalign{\kern-1pt\nointerlineskip}
 $\hfil\displaystyle{#1}\hfil$\crcr}}}

\newlength{\minitwocolumn}
\begin{document}

\begin{flushright}
DPUR/TH/57\\
November, 2017\\
\end{flushright}
\vspace{20pt}

\pagestyle{empty}
\baselineskip15pt

\begin{center}
{\large\bf Gravitational Quantum Effects and Nonlocal Approach to the Cosmological Constant Problem
\vskip 1mm }

\vspace{10mm}
Ichiro Oda \footnote{E-mail address:\ ioda@phys.u-ryukyu.ac.jp
}

\vspace{3mm}
           Department of Physics, Faculty of Science, University of the 
           Ryukyus,\\
           Nishihara, Okinawa 903-0213, Japan.\\

\end{center}


\vspace{3mm}
\begin{abstract}
We have recently presented a manifestly local and general coordinate invariant formulation of a nonlocal approach
to the cosmological constant problem. In this article, we investigate quantum effects from both matter and gravitational 
fields in this formulation. In particular, we pay our attention to the gravitational loop effects and show that 
the effective cosmological constant is radiatively stable even in the presence of the gravitational loop effects
in addition to matter loop effects. For this purpose we need to add the $R^2$ term and the corresponding topological action
as an total action, which should be contrasted with the work by Kaloper and Padilla where the topological Gauss-Bonnet
term is added instead of the $R^2$ term. The advantages of our new formulation compared to that by Kaloper 
and Padilla are that not only we do not have to assume the scale invariance which is required to render the space-time
average of the square of the Weyl curvature vanishing, but also our formulation would lead to the $R^2$ inflation
in a natural manner.  
\end{abstract}

\newpage
\pagestyle{plain}
\pagenumbering{arabic}


\rm
\section{Introduction}

It is often said that one of the most serious contradictions in modern physics is the enormous mismatch between the observed value 
of cosmological constant and estimate of the contributions of elementary particles to the vacuum energy density.
Originally, the cosmological constant was introduced by Einstein as an undetermined constant in the Einstein equation 
so that this equation has a static cosmological solution, guided by his prejudice at that time that our Universe is static. 
However, the obtained solution is unstable and is not realistic. Afterwards, it turned out that we live in a dynamically expanding 
universe as theoretically pointed out by Friedmann and observationally discovered by Hubble.  
   
The cosmological constant $\Lambda$ is equivalent to the vacuum energy density $\rho$ via a relation 
$\rho = \frac{\Lambda}{8 \pi G} \equiv M_{Pl}^2 \Lambda$ where $G$ is the Newton constant and $M_{Pl} \sim 10^{18}
GeV$ is the reduced Planck mass. In quantum field theory (QFT), the vacuum energy density is theoretically evaluated 
by summing up the zero-point fluctuation of a quantum field up to a momentum cutoff $k_c$:
\begin{eqnarray}
\rho_{th} = \int^{k_c} \frac{d^3 k}{(2 \pi)^3} \frac{1}{2} \sqrt{\vec{k}^2 + m^2} \sim \frac{k_c^4}{16 \pi^2}.
\label{rho}
\end{eqnarray}
If the cutoff is chosen to the Planck mass, $k_c \sim M_{Pl}$, the theoretical value of the vacuum energy density
is $\rho_{th} \sim M_{Pl}^4$.  However, the observed cosmic vacuum energy density is known to be    
$\rho_{obs} \sim 10^{-120} M_{Pl}^4 \sim 10^{-120} \rho_{th}$. On the other hand, if the cutoff is chosen to 
the electro-weak scale $k_c \sim 1 TeV$, then we have $\rho_{th} \sim (1 TeV)^4$ so we obtain $\rho_{obs} \sim
10^{-60} (1 TeV)^4 \sim 10^{-60} \rho_{th}$. Confronting such a huge discrepancy of 60-120 digits between the theoretical value 
and the observed one, it has been conjectured for a long time that there might exist some mechanism to make the cosmological
constant exactly vanish, and its quest is called the cosmological constant problem \cite{Weinberg, Padilla}.

In some respects, one of the biggest problems of the cosmological constant problem is that it is rarely stated properly. To solve it, 
we had better to be clear what the problem really is. In this section, in order to account for the cosmological constant problem clearly,
let us confine ourselves to the semiclassical approach where matter loops are considered while quantum gravity effects are 
ignored.  

As a simple example, let us consider a real scalar field of mass $m$ with $\lambda \phi^4$ interaction, which is minimally coupled to 
the classical gravity:
\begin{eqnarray}
S  = \int d^4 x \sqrt{- g} \ \Biggl[ \frac{M_{Pl}^2}{2} R - \Lambda_b - \frac{1}{2} g^{\mu\nu} \partial_\mu \phi \partial_\nu \phi
- \frac{m^2}{2} \phi^2 - \frac{\lambda}{4!} \phi^4  \Biggr],
\label{phi4}
\end{eqnarray}
where $\Lambda_b$ is the bare cosmological constant which is divergent.  Using the dimensional regularization, it is straightforward to 
calculate 1-loop contribution of the scalar field to the cosmological constant by evaluating the functional determinant:
\begin{eqnarray}
V^{\phi, 1-loop} &\equiv& \frac{i}{2} tr \left[ \log \left( - i \frac{\delta^2 S}{\delta \phi^2} \right) \right]
= \frac{1}{2} \int \frac{d^4 k_E}{(2 \pi)^4} \log ( k_E^2 + m^2 )
\nonumber\\
&=&  - \frac{m^4}{(8 \pi)^2} \left[ \frac{2}{\epsilon}  + \log \left( \frac{\mu^2}{m^2} \right) + finite \right],
\label{V-1 loop}
\end{eqnarray}
where $\mu$ is the renormalization mass scale. In order to cancel the divergence associated with a simple pole $\frac{2}{\epsilon}$,
we are required to take the bare cosmological constant at the 1-loop level
\begin{eqnarray}
\Lambda^{\phi, 1-loop}_b = \frac{m^4}{(8 \pi)^2} \left[ \frac{2}{\epsilon}  + \log \left( \frac{\mu^2}{M^2} \right) \right],
\label{Bare-Lambda-1 loop}
\end{eqnarray}
where $M$ is an arbitrary substraction mass scale where the measurement is carried out.  Then, by summing up the two contributions,
the 1-loop renormalized cosmological constant is obtained: 
\begin{eqnarray}
\Lambda^{\phi, 1-loop}_{ren} = V^{\phi, 1-loop} + \Lambda^{\phi, 1-loop}_b
= \frac{m^4}{(8 \pi)^2} \left[ \log \left( \frac{m^2}{M^2} \right) - finite \right].
\label{Renorm-Lambda-1 loop}
\end{eqnarray}
This 1-loop renormalized cosmological constant is finite, but depends on the arbitrary scale M, so we cannot have a concrete prediction. 
According to QFT, what we need to do is to replace it with the measured value, not predict the value theoretically. 
Once this is done, one can proceed further and make predictions about all the physical quantities which are not ultra-violet (UV) sensitive.

Cosmological observation requires us to take $\Lambda^{\phi, 1-loop}_{ren} \sim (1 meV)^4$. If the particle mass $m$ is chosen to 
the electro-weak scale, we have $V^{\phi, 1-loop} \sim (1 TeV)^4 = 10^{60} (1 meV)^4$. Thus, the measurement suggests that 
the finite contribution to the 1-loop renormalized cosmological constant is cancelled to an accuracy of one part in $10^{60}$ between
$V^{\phi, 1-loop}$ and $\Lambda^{\phi, 1-loop}_b$. This big fine tuning is sometimes called the cosmological constant problem as well.
Following the lore of QFT, at this stage of the argument, we have no issue with this fine tuning.

However, the issue arises when we go up to higher loops. For instance, at the 2-loop level, $V^{\phi, 2-loop}$ is proportional to $\lambda m^4$.
(In general, at the n-loop level,  $V^{\phi, n-loop} \propto \lambda^{n-1} m^4$.) Then, the consistency between the measurement and
the perturbation theory requires us to set up an equality
\begin{eqnarray}
(1 meV)^4 = \Lambda_{ren} = \Lambda_b + V^{\phi, 1-loop} + V^{\phi, 2-loop} + V^{\phi, 3-loop} + \cdots.
\label{Renorm-Lambda-n loop}
\end{eqnarray}
The problem is that each $V^{\phi, i-loop} (i = 1, 2, \cdots)$ has almost the same and huge size compared to the observed value
$(1 meV)^4$.  Thus, even if we fined tune the cosmological constant at the 1-loop level, the fine tuning is spoilt at the 2-loop level
so that we must retune the finite contribution in the bare cosmological constant term to the same degree of accuracy. In other words, 
at each successive order in perturbation theory, we are required to fine tune to extreme accuracy!  This problem is called "radiative instability", 
i.e., the need to repeatedly fine tune whenever the higher loop corrections are included, which is the essence of the cosmological constant
problem. What this is telling us is that the cosmological constant is very sensitive to the details of UV physics which we are 
ignorant in the effective field theory.

In order to solve the problem of the radiative instability of the cosmological constant, some nonlocal formulations have been advocated
\cite{Linde}-\cite{Oda3}, but many of them have been restricted to the semiclassical approach where only radiative corrections 
from matter fields are considered whereas the gravity is treated with as a classical field merely serving for detecting the vacuum energy. 
Recently, an interesting approach, which attempts to deal with the graviton loop effects by using the topological Gauss-Bonnet term, 
has been proposed \cite{Kaloper3}. 
In this formalism, it is necessary that the Weyl tensor does not receive any large scale contribution from the radiatively unstable vacuum
energy. Indeed, our universe can be nicely described by a spatially flat FLRW cosmology which is conformally flat, thereby making the
Weyl tensor be vanishing. 

In this article, we wish to present an alternative formalism attacking the same problem. It is well known that as a result of renormalization 
of matter and gravitational fields, the higher derivative terms, such as $R^2$ and $R_{\mu\nu}^2$, are naturally generated 
\cite{Visser, Birrell, Parker}, so that they should be handled on the same footing as the Gauss-Bonnet term. In our new formalism, 
we would like to construct a nonlocal approach to the cosmological constant problem
by incorporating such the higher-derivative terms in the framework of quantum gravity where radiative corrections from both matter and
gravitational fields are properly treated with.  As a bonus, our formalism would account for inflation universe $\acute{a}$ la Starobinsky 
\cite{Starob} where the $R^2$ term plays a critical role.

\section{Review of manifestly local formulation}

We start by reviewing a manifestly local and generally coordinate invariant formulation \cite{Oda1, Oda2} for a nonlocal approach 
to the cosmological constant problem.\footnote{See the related papers \cite{Oda4}-\cite{Oda6}.}  

A manifestly local and generally coordinate invariant action for our nonlocal approach takes the form 
\begin{eqnarray}
S = S_{GR} + S_{Top},
\label{T-Action}
\end{eqnarray}
where the gravitational action $S_{GR}$ with a generic matter Lagrangian density ${\cal{L}}_{m}$ and the
topological action $S_{Top}$ are respectively defined as \footnote{This action is also the hybrid action discussed in \cite{Kaloper4}.}
\begin{eqnarray}
S_{GR}  = \int d^4 x \sqrt{- g} \ \Biggl[ \eta(x) ( R - 2 \Lambda ) + {\cal{L}}_{m} 
- \frac{1}{2} \cdot \frac{1}{4!} F_{\mu\nu\rho\sigma}^2 
+ \frac{1}{6} \nabla_\mu ( F^{\mu\nu\rho\sigma} A_{\nu\rho\sigma} )
\Biggr].
\label{GR-Action}
\end{eqnarray}
and
\begin{eqnarray}
S_{Top}  = \int d^4 x  \ \frac{1}{4!}  \ \mathring{\varepsilon}^{\mu\nu\rho\sigma} M_{Pl}^2
f\left( \frac{\eta(x)}{M_{Pl}^2}  \right) H_{\mu\nu\rho\sigma}.
\label{Top-Action}
\end{eqnarray}

Since we have introduced various quantities in the above equations, we wish to account for their definitions
in what follows:\footnote{We follow notation and conventions of the textbook by Misner et al \cite{MTW}.} 
$g$ is the determinant of the metric tensor, $g = \det g_{\mu\nu}$, and $R$ denotes the
scalar curvature. $\eta(x)$ is a scalar field of dimension of mass squared. Let us note that this scalar
field  $\eta(x)$ has no local degrees of freedom except the zero mode because of the gauge symmetry
of the 4-form strength \cite{Henneaux}. $\Lambda$ and $ {\cal{L}}_{m}$ 
are the bare cosmological constant and the Lagrangian density for generic matter fields, respectively.  
Moreover, $F_{\mu\nu\rho\sigma}$ and $ H_{\mu\nu\rho\sigma}$ are respectively the field strengths 
for two 3-form gauge fields $A_{\mu\nu\rho}$ and $B_{\mu\nu\rho}$
\begin{eqnarray}
F_{\mu\nu\rho\sigma} = 4 \partial_{[\mu} A_{\nu\rho\sigma]},  \qquad 
H_{\mu\nu\rho\sigma} = 4 \partial_{[\mu} B_{\nu\rho\sigma]},
\label{4-forms}
\end{eqnarray}
where the square brackets denote antisymmetrization of enclosed indices. Finally, $\mathring{\varepsilon}^{\mu\nu\rho\sigma}$
and $\mathring{\varepsilon}_{\mu\nu\rho\sigma}$ are the Levi-Civita tensor density defined as
\begin{eqnarray}
\mathring{\varepsilon}^{0123} = + 1,  \qquad 
\mathring{\varepsilon}_{0123} = - 1,
\label{Levi}
\end{eqnarray}
and they are related to the totally antisymmetric tensors $\varepsilon^{\mu\nu\rho\sigma}$ and $\varepsilon_{\mu\nu\rho\sigma}$ via
relations
\begin{eqnarray}
\varepsilon^{\mu\nu\rho\sigma} = \frac{1}{\sqrt{-g}}  \mathring{\varepsilon}^{\mu\nu\rho\sigma},  \qquad 
\varepsilon_{\mu\nu\rho\sigma} = \sqrt{-g}  \mathring{\varepsilon}_{\mu\nu\rho\sigma}.
\label{Levi-tensor}
\end{eqnarray}
Also note that the Levi-Civita tensor density satisfies the following equations:
\begin{eqnarray}
\mathring{\varepsilon}^{\mu\nu\rho\sigma} \mathring{\varepsilon}_{\alpha\beta\rho\sigma} = - 2 ( \delta_\alpha^\mu \delta_\beta^\nu
-   \delta_\alpha^\nu \delta_\beta^\mu ),  \qquad 
\mathring{\varepsilon}^{\mu\nu\rho\sigma} \mathring{\varepsilon}_{\alpha\nu\rho\sigma} = - 3!  \delta_\alpha^\mu,  \qquad 
\mathring{\varepsilon}^{\mu\nu\rho\sigma} \mathring{\varepsilon}_{\mu\nu\rho\sigma} = - 4!.
\label{e-identity}
\end{eqnarray}
Finally, we have introduced a smooth function $f(x)$ which cannot be a linear function.  

Now let us derive all the equations of motion from the action (\ref{T-Action}).  First of all, the variation 
with respect to the 3-form $B_{\mu\nu\rho}$ gives rise to the equation for a scalar field  $\eta(x)$:
\begin{eqnarray}
\mathring{\varepsilon}^{\mu\nu\rho\sigma} f^\prime \partial_\sigma \eta(x)  = 0,
\label{B-eq}
\end{eqnarray}
where $f^\prime (x) \equiv \frac{d f(x)}{dx}$.  From this equation, we have a classical solution for $\eta(x)$,
\begin{eqnarray}
\eta(x)  = \eta,
\label{eta-sol}
\end{eqnarray}
where $\eta$ is a certain constant.  Next, taking the variation of the scalar field $\eta(x)$ leads to the
equation:
\begin{eqnarray}
\sqrt{- g} ( R - 2 \Lambda ) + \frac{1}{4!}  \ \mathring{\varepsilon}^{\mu\nu\rho\sigma} f^\prime H_{\mu\nu\rho\sigma} = 0.
\label{eta-equation}
\end{eqnarray}
Since we can always set $H_{\mu\nu\rho\sigma}$ to be
\begin{eqnarray}
H_{\mu\nu\rho\sigma}  = c(x) \varepsilon_{\mu\nu\rho\sigma} = c(x) \sqrt{-g} \mathring{\varepsilon}_{\mu\nu\rho\sigma},
\label{H-equation}
\end{eqnarray}
with $c(x)$ being some scalar function, using the last equation in Eq. (\ref{e-identity}), Eq.  (\ref{H-equation}) can be rewritten as
\begin{eqnarray}
R - 2 \Lambda - f^\prime c(x) = 0.
\label{H-equation 2}
\end{eqnarray}
In order to take account of the cosmological constant problem, let us take the space-time average of this equation,
which gives us a constraint equation:
\begin{eqnarray}
\overline{R} = 2 \Lambda + f^\prime \left( \frac{\eta}{M_{Pl}^2} \right) \overline{c(x)},
\label{H-constraint}
\end{eqnarray}
where we have used Eq. (\ref{eta-sol}), and for a generic space-time dependent quantity $Q(x)$, the space-time 
average is defined as 
\begin{eqnarray}
\overline{Q(x)}  = \frac{\int d^4 x \sqrt{-g} \, Q(x)}{\int d^4 x \sqrt{-g}},
\label{ST average}
\end{eqnarray}
where the denominator $V \equiv \int d^4 x \sqrt{-g}$ denotes the space-time volume.   

The equation of motion for the 3-form $A_{\mu\nu\rho}$ gives the Maxwell-like equation
\begin{eqnarray}
\nabla^\mu F_{\mu\nu\rho\sigma} = 0.
\label{A-eq}
\end{eqnarray}
As in $H_{\mu\nu\rho\sigma}$, if we set 
\begin{eqnarray}
F_{\mu\nu\rho\sigma}  = \theta(x) \varepsilon_{\mu\nu\rho\sigma} = \theta(x) \sqrt{-g} 
\mathring{\varepsilon}_{\mu\nu\rho\sigma},
\label{F}
\end{eqnarray}
with $\theta(x)$ being a scalar function, Eq. (\ref{A-eq}) requires $\theta(x)$ to be a constant
\begin{eqnarray}
\theta(x) = \theta,
\label{Theta}
\end{eqnarray}
where $\theta$ is a constant. 

Finally, the variation with respect to the metric tensor yields the gravitational field equation, i.e., the Einstein equation:
\begin{eqnarray}
\eta \left(  G_{\mu\nu}  + \Lambda g_{\mu\nu} \right)  - \frac{1}{2} T_{\mu\nu}
+ \frac{1}{4} \cdot \frac{1}{4!} g_{\mu\nu} F_{\alpha\beta\gamma\delta}^2 
- \frac{1}{12} F_{\mu\alpha\beta\gamma} F_\nu \,^{\alpha\beta\gamma}
= 0,
\label{Eins-eq 1}
\end{eqnarray}
where $G_{\mu\nu} = R_{\mu\nu} - \frac{1}{2} g_{\mu\nu} R$ is the well-known Einstein tensor and
the energy-momentum tensor is defined by $ T_{\mu\nu} = - \frac{2}{\sqrt{-g}} \frac{\delta (\sqrt{-g}  {\cal{L}}_{m})}
{\delta g^{\mu\nu}}$ as usual.  In deriving Eq. (\ref{Eins-eq 1}), we have used Eq. (\ref{eta-sol}).
Furthermore, using Eqs. (\ref{F}) and (\ref{Theta}), this equation can be simplified to be the form
\begin{eqnarray}
\eta \left(  G_{\mu\nu}  + \Lambda g_{\mu\nu} \right)  - \frac{1}{2} T_{\mu\nu}
+ \frac{1}{4} \theta^2 g_{\mu\nu} = 0.
\label{Eins-eq 1-theta}
\end{eqnarray}

Taking the trace of Eq.  (\ref{Eins-eq 1-theta}) and then the space-time average, one obtains a constraint
\begin{eqnarray}
\theta^2 = \frac{1}{2} \overline{T} + \eta ( \overline{R} - 4 \Lambda).
\label{Eins-constraint}
\end{eqnarray}
Substituting this constraint into the Einstein equation (\ref{Eins-eq 1-theta}), we find that
\begin{eqnarray}
M_{Pl}^2 G_{\mu\nu}  + \frac{1}{4} M_{Pl}^2 \overline{R}  g_{\mu\nu} = T_{\mu\nu}
- \frac{1}{4} \overline{T} g_{\mu\nu},
\label{Final Eins-eq 1}
\end{eqnarray}
where we have chosen $\eta = \frac{M_{Pl}^2}{2}$. Next, let us separate the energy-momentum tensor
$T_{\mu\nu}$ into two parts
\begin{eqnarray}
T_{\mu\nu} = - V_{vac} g_{\mu\nu} + \tau_{\mu\nu},
\label{Energy-momentum}
\end{eqnarray}
where $V_{vac}$ denotes the sum of a classical vacuum energy and a quantum vacuum correction
coming from matter fields to an arbitrary order in the loop expansion, and $\tau_{\mu\nu}$ is
the local excitation such as radiation, which is a finite quantity. Inserting Eq. (\ref{Energy-momentum})
to Eq. (\ref{Final Eins-eq 1}), we arrive at the desired Einstein equation
\begin{eqnarray}
M_{Pl}^2 G_{\mu\nu}  + \frac{1}{4} M_{Pl}^2 \overline{R}  g_{\mu\nu} = \tau_{\mu\nu}
- \frac{1}{4} \overline{\tau} g_{\mu\nu}.
\label{Final Eins-eq 1-1}
\end{eqnarray}
  
The Einstein equation (\ref{Final Eins-eq 1-1}) shows that the vacuum energy $V_{vac}$ decouples
from the gravitational field equation and gives us information on the effective cosmological constant
\begin{eqnarray}
\Lambda_{eff} =  \frac{1}{4} M_{Pl}^2 \overline{R}  + \frac{1}{4} \overline{\tau}.
\label{CC-1}
\end{eqnarray}
The first term in the right-handed side (RHS) is radiatively stable since $\overline{R}$ is so.
Actually, as seen in Eq.  (\ref{H-constraint}), $\Lambda$ is a mere number and $\overline{c(x)}$
is proportional to the flux of the 4-form which is the IR quantity. The second term in the RHS
is obviously radiatively stable. Thus, our cosmological constant $\Lambda_{eff}$ is a radiatively
stable quantity so it can be fixed by the measurement in a consistent manner.
 
The only disadvantage of our nonlocal approach to the cosmological constant problem is that
we confine ourselves to the semiclassical approach where the matter loop effects are included
in the energy momentum tensor while the quantum gravity effects are completely ignored
(Gravity is a classical field merely serving the purpose of detecting the vacuum energy).
In the next section, we will take the quantum gravity effects into consideration.

\section{Quantum gravity effects}

From the 1-loop calculation, the dimensional analysis and general covariance, it is easy to estimate the loop 
effects from both matter and the gravitational fields.  For instance, the renormalization of the Newton's constant 
and the cosmological constant amounts to adding the following action to the total action (\ref{T-Action})
up to the logarithmic divergences which are irrelevant to the argument at hand:
\begin{eqnarray}
S_q  &=& \int d^4 x \sqrt{- g} \ \Biggl[ \left( a_0 M^2 + a_1 \frac{M^4}{\eta} + a_2 \frac{M^6}{\eta^2}
+ \cdots \right) R + b_0 M^4 + b_1 \frac{M^6}{\eta} + b_2 \frac{M^8}{\eta^2} + \cdots \Biggr]
\nonumber\\
&\equiv& \int d^4 x \sqrt{- g} \ \left[ \alpha (\eta) R + \beta(\eta) \right],
\label{q-Action}
\end{eqnarray}
where $M$ is a cutoff and the coefficients $a_i, b_i ( i = 0, 1, 2, \cdots )$ are ${\cal{O}}(1)$.  

In addition to it, quantum effects from the matter and gravitational fields lead to the higher-derivative terms
\begin{eqnarray}
S_h = \int d^4 x \sqrt{- g} ( k_1 R^2 + k_2 R_{\mu\nu}^2 + k_3 R_{\mu\nu\rho\sigma}^2 ),
\label{High-deriv-Action}
\end{eqnarray}
where $k_i (i = 1, 2, 3)$ are some constants and only two among three terms in the RHS are independent 
owing to the fact that in four space-time dimensions, 
\begin{eqnarray}
\int d^4 x \sqrt{-g} E 
\equiv \int d^4 x \sqrt{-g} ( R_{\mu\nu\rho\sigma}^2 - 4 R_{\mu\nu}^2 + R^2 ),
\label{Euler}
\end{eqnarray}
is a topological invariant called the Euler number. 
In this section, the higher-derivative terms in Eq. (\ref{High-deriv-Action}) are ignored
and they will be treated with in the next section.

Now the total action is given by
\begin{eqnarray}
S = S_{GR} + S_{Top} + S_q,
\label{T-Action 2}
\end{eqnarray}
Compared to the case in the previous section, only the modification of field equations lies in field 
equations with respect to the scalar field $\eta(x)$ and the metric tensor $g_{\mu\nu}$.
First, the field equation for $\eta(x)$ reads
\begin{eqnarray}
R = \frac{1}{1 + \alpha^\prime(\eta)}  \left[ 2 \Lambda - \beta^\prime(\eta) 
+ f^\prime \left( \frac{\eta}{M_{Pl}^2} \right) c(x) \right].
\label{H-equation 2-1}
\end{eqnarray}
Consequently, the constraint (\ref{H-constraint}) is modified to be
\begin{eqnarray}
\overline{R} = \frac{1}{1 + \alpha^\prime(\eta)}  \left[ 2 \Lambda - \beta^\prime(\eta) 
+ f^\prime \left( \frac{\eta}{M_{Pl}^2} \right) \overline{c(x)} \right],
\label{H-constraint 2}
\end{eqnarray}
Next, the gravitational equation turns out to be changed to 
\begin{eqnarray}
M_{eff}^2 G_{\mu\nu}  + \frac{1}{4} M_{eff}^2 \overline{R}  g_{\mu\nu} = \tau_{\mu\nu}
- \frac{1}{4} \overline{\tau} g_{\mu\nu},
\label{Final Eins-eq 2}
\end{eqnarray}
where $M_{eff}^2 \equiv 2 (\eta + \alpha(\eta))$.  This equation shows that the effective cosmological 
constant takes the form
\begin{eqnarray}
\Lambda_{eff} =  \frac{1}{4} M_{eff}^2 \overline{R}  + \frac{1}{4} \overline{\tau}.
\label{CC-2}
\end{eqnarray}
The second term in the RHS is obviously radiatively stable. However, it is not clear that the first term in the RHS
is radiatively stable or not since $\overline{R}$ is now given by Eq. (\ref{H-constraint 2}).

For the clarity of the argument, suppose that the cutoff $M$ is the GUT scale $M \sim M_{GUT} = 10^{16} GeV$
and $\eta \sim M_{Pl} = 10^{18} GeV$.  $\alpha^\prime(\eta)$ and $\beta^\prime(\eta)$ appearing in 
Eq. (\ref{H-constraint 2}) are calculated to be
\begin{eqnarray}
\alpha^\prime(\eta) &=& - a_1 \frac{M^4}{\eta^2} - 2 a_2 \frac{M^6}{\eta^3} - 3 a_3 \frac{M^8}{\eta^4}
- \cdots       \nonumber\\
\beta^\prime(\eta) &=& - b_1 \frac{M^6}{\eta^2} - 2 b_2 \frac{M^8}{\eta^3} - 3 b_3 \frac{M^{10}}{\eta^4}
- \cdots. 
\label{alpha-beta}
\end{eqnarray}
Since $\frac{M^4}{\eta^2} \sim 10^{-8}, \frac{M^6}{\eta^3} \sim 10^{-12}$ etc., the denominator in 
Eq. (\ref{H-constraint 2}) can be rewritten as
$1 + \alpha^\prime(\eta) \approx 1$ so it is radiatively stable. However, the part including
$\beta^\prime(\eta)$ in the numerator,
is found to be radiatively unstable by the following reasoning: 
\begin{eqnarray}
\Lambda_{eff} &\sim& M_{Pl}^2 \overline{R} \sim M_{Pl}^2 \beta^\prime(\eta)
\nonumber\\
&\sim& M_{Pl}^4 \left(  - b_1 \cdot 10^{-12} - b_2 \cdot 10^{-16} - b_3 \cdot 10^{-20} - \cdots \right)
\nonumber\\
&\sim& 1 (meV)^4.
\label{CC-2-1}
\end{eqnarray}
At the last step, we used the cosmic observed value of the cosmological constant. Each term in the RHS has 
a much larger value compared to $1 (meV)^4$, which implies that  $\beta^\prime(\eta)$ is not radiatively stable. 
Thus, we can conclude that the effective cosmological constant $\Lambda_{eff}$ is not radiatively 
stable in this case.

\section{$R^2$-gravity model}

As seen in the previous section, the effective cosmological constant $\Lambda_{eff}$ is not radiatively stable
when graviton loop effects are incorporated into the nonlocal approach to the cosmological constant problem.
To remedy this situation, it is natural to consider the remaining quantum effect, that is, the higher-derivative
terms  (\ref{High-deriv-Action}). Since we expect that the present theory can also provide us with the inflation
universe, let us first consider the case of the $R^2$ term. The case of $R_{\mu\nu}^2$ will be considered afterwards.

Our total action is, therefore, constituted of four sectors:
\begin{eqnarray}
S = S_{GR} + S_{Top} + S_q + S_{R^2},
\label{T-R2-Action}
\end{eqnarray}
where the last action $S_{R^2}$ is defined as
\begin{eqnarray}
S_{R^2}  = \int d^4 x \sqrt{- g} \omega(x) R^2 + \int d^4 x  \ \frac{1}{4!}  \ \mathring{\varepsilon}^{\mu\nu\rho\sigma} 
M_{Pl}^2 \hat f(\omega) \hat H_{\mu\nu\rho\sigma},
\label{R2-Action}
\end{eqnarray}
where $\hat H_{\mu\nu\rho\sigma} \equiv 4 \partial_{[\mu} \hat B_{\nu\rho\sigma]}$. We might be concerned that 
that the higher-derivative term $\omega R^2$ would generate new radiative corrections that also 
depend on $\omega$, thereby inducing new radiative corrections to the vacuum energy density and consequently breaking 
its radiative stability. However, since the $R^2$ term is a renormalizable term, the radiative corrections are logarithmically 
divergent quantities \cite{Buchbinder}. It is straightforward to show that this is not indeed the case explicitly when the mass 
of the scalaron is around $1 meV$. The detail will be explained in a separate publication when we treat with the cosmic acceleration
on the basis of the present formalism \cite{Oda7}. 
 
The variation of the total action with respect to the 3-form $\hat B_{\mu\nu\rho}$ produces
\begin{eqnarray}
\mathring{\varepsilon}^{\mu\nu\rho\sigma} \hat f^\prime \partial_\sigma \omega(x)  = 0,
\label{hat-B-eq}
\end{eqnarray}
which gives us a classical solution for $\omega(x)$,
\begin{eqnarray}
\omega(x) = \omega,
\label{omega-sol}
\end{eqnarray}
where $\omega$ is a certain constant.  Next, taking the variation of the scalar field $\omega(x)$ yields the
field equation:
\begin{eqnarray}
\sqrt{- g} R^2 + \frac{1}{4!}  \ \mathring{\varepsilon}^{\mu\nu\rho\sigma} M_{Pl}^2 \hat f^\prime 
\hat H_{\mu\nu\rho\sigma} = 0.
\label{omega-equation}
\end{eqnarray}
Setting $\hat H_{\mu\nu\rho\sigma}$ again to be
\begin{eqnarray}
\hat H_{\mu\nu\rho\sigma}  = \hat c(x) \varepsilon_{\mu\nu\rho\sigma} 
= \hat c(x) \sqrt{-g} \mathring{\varepsilon}_{\mu\nu\rho\sigma},
\label{hat-H-equation}
\end{eqnarray}
with $\hat c(x)$ being some scalar function, Eq.  (\ref{omega-equation}) can be cast to
\begin{eqnarray}
R^2 - M_{Pl}^2 \hat f^\prime \hat c(x) = 0.
\label{hat-H-equation 2}
\end{eqnarray}
Then, the space-time average of this equation leads to a new constraint equation:
\begin{eqnarray}
\overline{R^2} = M_{Pl}^2 \hat f^\prime (\omega) \overline{\hat c(x)}.
\label{hat-H-constraint}
\end{eqnarray}
Note that $\overline{R^2}$ is radiatively stable since both $\hat f^\prime (\omega)$ and $ \overline{\hat c(x)}$
are radiatively stable. At this stage, it is worthwhile to comment on one remark: The scalar curvature in Eq. (\ref{H-equation 2-1})
is not equivalent to that in Eq.  (\ref{hat-H-equation 2}) since the latter scalar curvature includes the contribution
from the higher-derivative term and is given in Eq. (\ref{Tr-Eins-eq 3-3}) as seen shortly.

With the help of Eqs. (\ref{eta-sol}),  (\ref{F}), (\ref{Theta}) and (\ref{omega-sol}), the variation with respect to 
the metric tensor yields the Einstein equation:
\begin{eqnarray}
\eta \left(  G_{\mu\nu}  + \Lambda g_{\mu\nu} \right)  + \omega \ {}^{(1)} H_{\mu\nu} 
+ \alpha(\eta) G_{\mu\nu} - \frac{1}{2} \beta(\eta) g_{\mu\nu} - \frac{1}{2} T_{\mu\nu}
+ \frac{1}{4} \theta^2 g_{\mu\nu} = 0,
\label{Grav-theta}
\end{eqnarray}
where ${}^{(1)} H_{\mu\nu}$ is defined as
\begin{eqnarray}
{}^{(1)} H_{\mu\nu} &=&  \frac{1}{\sqrt{-g}} \frac{\delta}{\delta g^{\mu\nu}} \int d^n x \sqrt{-g} R^2
\nonumber\\
&=& -2 \nabla_\mu \nabla_\nu R + 2 g_{\mu\nu} \Box R - \frac{1}{2} g_{\mu\nu} R^2 + 2 R R_{\mu\nu}.
\label{Def H1}
\end{eqnarray}
with $\Box$ being a covariant d'Alembertian operator $\Box = g^{\mu\nu} \nabla_\mu \nabla_\nu$.
An important property of this tensor is 
\begin{eqnarray}
\overline{{}^{(1)} H^\mu _\mu} = 6 \overline{\Box R} = 0,
\label{Trace-H1}
\end{eqnarray}
up to a surface term which is now neglected. Following the same line of the argument as before, it is easy to 
arrive at the final form of the Einstein equation
\begin{eqnarray}
M_{eff}^2 G_{\mu\nu}  + 2 \omega \ {}^{(1)} H_{\mu\nu} + \frac{1}{4} M_{eff}^2 \overline{R}  g_{\mu\nu} 
= \tau_{\mu\nu} - \frac{1}{4} \overline{\tau} g_{\mu\nu},
\label{Eins-eq 3}
\end{eqnarray}
which indicates that the effective cosmological constant is again of the form
\begin{eqnarray}
\Lambda_{eff} =  \frac{1}{4} M_{eff}^2 \overline{R}  + \frac{1}{4} \overline{\tau}.
\label{CC-3}
\end{eqnarray}

Since the second term in the RHS is obviously radiatively stable, let us focus our attention on the first term:
\begin{eqnarray}
\Delta \Lambda \equiv  \frac{1}{4} M_{eff}^2 \overline{R}.
\label{CC-4}
\end{eqnarray}
Then, the Einstein equation (\ref{Eins-eq 3}) reads
\begin{eqnarray}
M_{eff}^2 G_{\mu\nu}  + 2 \omega \ {}^{(1)} H_{\mu\nu} +  \Delta \Lambda g_{\mu\nu} 
= \tau_{\mu\nu} - \frac{1}{4} \overline{\tau} g_{\mu\nu}.
\label{Eins-eq 3-2}
\end{eqnarray}
Taking the trace of this equation, one obtains:
\begin{eqnarray}
M_{eff}^2 \left( 1 - \frac{12 \omega}{M_{eff}^2} \Box \right) R  
= 4 \left[ \Delta \Lambda - \frac{1}{4} ( \tau - \overline{\tau} ) \right],
\label{Tr-Eins-eq 3-2}
\end{eqnarray}
from which one can express the scalar curvature as
\begin{eqnarray}
R = \frac{4}{M_{eff}^2} \left[ \Delta \Lambda - \frac{\tau - \overline{\tau}}{4} 
+ \frac{1}{4} \frac{\Box}{\Box - \frac{M_{eff}^2}{12 \omega}}  \tau \right].
\label{Tr-Eins-eq 3-3}
\end{eqnarray}
Note that this expression reduces to Eq. (\ref{CC-4}) up to a surface term when one takes the space-time average,
which guarantees the correctness of our derivation. 

Taking the square of Eq. (\ref{Tr-Eins-eq 3-3}) and then the space-time average makes it possible to
describe $(\Delta \Lambda)^2$ in terms of $\overline{R^2}$ and $\tau$:
\begin{eqnarray}
(\Delta \Lambda)^2 =  \frac{M_{eff}^4}{16} \overline{R^2} - \frac{1}{16} \left( \overline{\tau^2} 
- \overline{\tau}^2 \right) - \frac{1}{16} \overline{\left( \frac{\Box}{\Box - \frac{M_{eff}^2}{12 \omega}}  \tau
\right)^2 } 
+ \frac{1}{8} \overline{\left(\tau \frac{\Box}{\Box - \frac{M_{eff}^2}{12 \omega}} \tau \right)}.
\label{Delta Lambda}
\end{eqnarray}
This expression clearly shows that $\Delta \Lambda$ is radiatively stable since $M_{eff}^2 \equiv 2 (\eta + \alpha(\eta))$
turns out to be radiatively stable, $ \overline{R^2}$ is also radiatively stable as shown in Eq. (\ref{hat-H-constraint}),
and the remaining terms involving $\tau$ are radiatively stable as well as long as $\Box$ is not equal to 
$\frac{M_{eff}^2}{12 \omega}$.\footnote{This is not a fine tuning.}  Incidentally, in the high energy limit, 
$(\Delta \Lambda)^2$ reduces to
\begin{eqnarray}
(\Delta \Lambda)^2 \rightarrow  \frac{M_{eff}^4}{16} \overline{R^2} + \frac{1}{16} \overline{\tau}^2,
\label{Delta Lambda 2}
\end{eqnarray}
which is manifestly radiatively stable. The radiative stability of $\Delta \Lambda$ ensures that the effective
cosmological constant $\Lambda_{eff}$ in Eq. (\ref{CC-3}) is indeed radiatively stable.

In this way, we have succeeded in proving that the effective cosmological constant is radiatively stable
when we add the action $S_{R^2}$ to the action of the previous section. A natural question is 
what becomes of the effective cosmological constant when we add the $R_{\mu\nu}^2$ term instead of the
$R^2$ term. The calculation proceeds in a perfectly similar manner to the case of the $R^2$ term. The
first modification is that the constraint (\ref{hat-H-constraint}) is replaced by a new constraint
\begin{eqnarray}
\overline{R_{\mu\nu}^2} = M_{Pl}^2 \hat f^\prime (\omega) \overline{\hat c(x)}.
\label{hat-H-constraint 2}
\end{eqnarray}
The second modification is to replace ${}^{(1)} H_{\mu\nu}$ with ${}^{(2)} H_{\mu\nu}$ in the Einstein equation
 (\ref{Eins-eq 3-2}) where ${}^{(2)} H_{\mu\nu}$ is defined as
\begin{eqnarray}
{}^{(2)} H_{\mu\nu} &=&  \frac{1}{\sqrt{-g}} \frac{\delta}{\delta g^{\mu\nu}} \int d^n x \sqrt{-g} R_{\alpha\beta}^2
\nonumber\\
&=& -2 \nabla_\rho \nabla_\mu R_\nu \, ^\rho + \Box R_{\mu\nu} + \frac{1}{2} g_{\mu\nu} \Box R 
+ 2 R_{\mu\rho} R_\nu \, ^\rho - \frac{1}{2} g_{\mu\nu} R_{\alpha\beta}^2.
\label{Def H2}
\end{eqnarray}
This tensor also satisfies an important property 
\begin{eqnarray}
\overline{{}^{(2)} H^\mu _\mu} = - 2 \overline{\nabla_\mu \nabla_\nu R^{\mu\nu}} + 3 \overline{\Box R} = 0,
\label{Trace-H2}
\end{eqnarray}
up to a surface term. 
 
As a result, the equation corresponding to Eq.  (\ref{Tr-Eins-eq 3-2}) now takes the form
\begin{eqnarray}
M_{eff}^2 \left[ \left( 1 - \frac{6 \omega}{M_{eff}^2} \Box \right) g^{\mu\nu} 
+ \frac{4 \omega}{M_{eff}^2} \nabla^\mu \nabla^\nu \right]  R_{\mu\nu}  
= 4 \left[ \Delta \Lambda - \frac{1}{4} ( \tau - \overline{\tau} ) \right].
\label{Tr-Eins-eq 4}
\end{eqnarray}
Owing to the tensorial character of an operator in front of $R_{\mu\nu}$, it seems to be difficult
to invert the operator to express $R_{\mu\nu}$ in terms of $\Delta \Lambda$ and $\tau$ 
in an analytical way. However, Eq. (\ref{Tr-Eins-eq 4})
implicitly shows that $R_{\mu\nu}$ could be expressed in terms of $\Delta \Lambda$ and the radiatively
stable $\tau$, so using the radiatively stable equation (\ref{hat-H-constraint 2}), $\Delta \Lambda$
could be described by only radiatively stable objects $\overline{R_{\mu\nu}^2}$ and $\tau$.
This would imply that the effective cosmological constant is also radiatively stable in the case of
the $R_{\mu\nu}^2$ term as in the $R^2$ case.

\section{Discussions}

In this article, we have investigated a possibility of incorporating the gravitational loop effects in the
framework of the nonlocal approach to the cosmological constant problem. To do that, we have added
the higher-derivative terms, which are always induced in a curved space-time via quantum effects, 
in addition to renormalization effects of the Newton constant and cosmological constant. In particular, 
in the case of the $R^2$ term, the analytical expression of the effective cosmological constant can be obtained 
by solving the Einstein equation with the help of a constraint equation connecting the space-time average of 
the $R^2$ term with the flux of the 4-form field strength. Consequently, it is explicitly found that 
the effective cosmological constant is radiatively stable even in the presence of the gravitational loop effects.

It is known that one subtle point, i.e., violation of unitarity, arises when there are the higher-derivative terms 
in an action. The present formalism, however, belongs to a category of the effective field theory, which is obtained
by integrating over the higher energy modes of an underlying UV-complete theory. Since our theory is not a UV-complete
theory, we are free of the issue of the unitarity violation owing to a massive ghost having the Planck mass.
An important future problem would be to construct an UV-complete theory corresponding to our formalism 
within the framework of string theory.  

As a bonus, we would have an opportunity to explain the inflation cosmology on the basis of the present
formulation since our model naturally includes the $R^2$ term triggering inflation as discovered 
by Starobinsky. It is worthwhile to point out that constructing of a model accounting for the inflation in 
the present context is a nontrivial task since the vacuum energy density decouples from the gravitational
equation and the residual cosmological constant is in general believed to be small in large and/or old 
universes. In a future publication, we will report this study \cite{Oda7}.

\begin{flushleft}
{\bf Acknowledgements}
\end{flushleft}

This work was supported by JSPS KAKENHI Grant Number 16K05327.
We are grateful for the warm hospitality of Dipartimento di Fisica e Astronomia "Galileo Galilei", 
Universita degli Studi di Padova, where we have initiated this work.


\end{document}